\documentclass[aps,prl,amsmath,amssymb,twocolumn,superscriptaddress,letterpaper]{revtex4}
\usepackage{dcolumn}
\usepackage{graphicx}
\usepackage{epstopdf}
\usepackage{verbatim}
\usepackage{amssymb}
\usepackage{amsmath}
\usepackage{subfigure,wrapfig}
\usepackage{color}

\begin{document}

\title{Imaging of Iso-frequency Contours via Resonance-Enhanced Scattering in Near-Pristine Photonic Crystals}

\author{Emma C. Regan$^{1,2\star}$, Yuichi Igarashi$^{1,3\star}$,
Bo Zhen$^{1,4\star}$, Ido Kaminer$^{1}$,\\ Chia Wei Hsu$^{5}$,  Yichen Shen$^{1}$, John D. Joannopoulos$^{1}$, and Marin Solja\v{c}i\'{c}}
\affiliation{
\normalsize{Research Laboratory of Electronics, Massachusetts Institute of Technology,}\\
\normalsize{Cambridge, MA 02139, USA}\\
\normalsize{$^{2}$Department of Physics, Wellesley College,}\\
\normalsize{Wellesley, MA 02481, USA}\\
\normalsize{$^{3}$Smart Energy Research Laboratories, NEC Corporation,}\\ 
\normalsize{34 Miyuiga-ka, Tsukuba, Ibaraki 305-8501, Japan}\\
\normalsize{$^{4}$Physics Department and Solid State Institute, Technion,}\\
\normalsize{Haifa 32000, Israel}\\
\normalsize{$^{5}$Department of Applied Physics, Yale University,} \\ 
\normalsize{New Haven, CT 06520, USA}\\
\normalsize{$^\star$These authors contributed equally to this work.} \\
\normalsize{Corresponding author: bozhen@mit.edu} 
}

\begin{abstract}

The iso-frequency contours of a photonic crystal are important for predicting and understanding exotic optical phenomena that are not apparent from high-symmetry band structure visualizations. Here, we demonstrate a method to directly visualize the iso-frequency contours of high-quality photonic crystal slabs that shows quantitatively good agreement with numerical results throughout the visible spectrum. Our technique relies on resonance-enhanced photon scattering from generic fabrication disorder and surface roughness, so it can be applied to general photonic and plasmonic crystals, or even quasi-crystals. We also present an analytical model of the scattering process, which explains the observation of iso-frequency contours in our technique. Furthermore, the iso-frequency contours provide information about the characteristics of the disorder and therefore serve as a feedback tool to improve fabrication processes.

\end{abstract}

\maketitle

While band structures are useful for understanding and predicting the optical properties of photonic crystal (PhC) slabs, they are typically calculated or measured only along high-symmetry directions because of computation and time constraints.  A more comprehensive understanding of a PhC slab lies in its iso-frequency contours: slices in 2D momentum space $(k_\text{x},k_\text{y})$ of constant frequency $\omega$. In particular, iso-frequency contours are essential for understanding phenomena that depend on the direction of group velocities, such as negative refraction \cite{PhysRevB.65.201104}, super-collimation \cite{:/content/aip/journal/apl/74/9/10.1063/1.123502,Rakichsuper} and super-lensing \cite{PhysRevB.58.R10096}. However, obtaining iso-frequency contours is not trivial, and current experimental techniques involve low-quality samples with strong disorder \cite{:/content/aip/journal/apl/97/25/10.1063/1.3524520}, additional fabrication steps \cite{PhysRevB.79.033305}, or sophisticated experimental setups \cite{Sapienza2012}. 

As an alternative, we demonstrate direct imaging of iso-frequency contours using resonance-enhanced photon scattering arising from minimal, intrinsic fabrication disorder in a "pristine" sample. The experimental iso-frequency contours show good quantitative agreement with numerical results throughout the visible wavelength regime. To understand the underlying physical process, we use temporal coupled mode theory \cite{joannopoulos2011photonic,Fan:03,Suh2004} to show that scattering is enhanced for on-resonance photons and that the angular distribution of scattered photons recreates the iso-frequency contours in the far field. Resonance-enhanced scattering has been used to study localized photonic modes \cite{McCutcheon2005,Galli2009,Portalupi2011}, but we show that this scattering process can also be used to probe large-area, delocalized resonances and the corresponding iso-frequency contours.

Additionally, our results show features that reflect information about the characteristic fabrication disorder in a PhC slab and thus serve as a feedback method for improving the fabrication processes. This technique may also be extended to quasi-crystals and provide an experimental route to obtain their iso-frequency contours, which are quite challenging to obtain using current numerical tools.    

\section*{Direct Imaging of Iso-frequency Contours} 

\begin{figure}[t!]
\begin{center}
\includegraphics[width=3.25in]{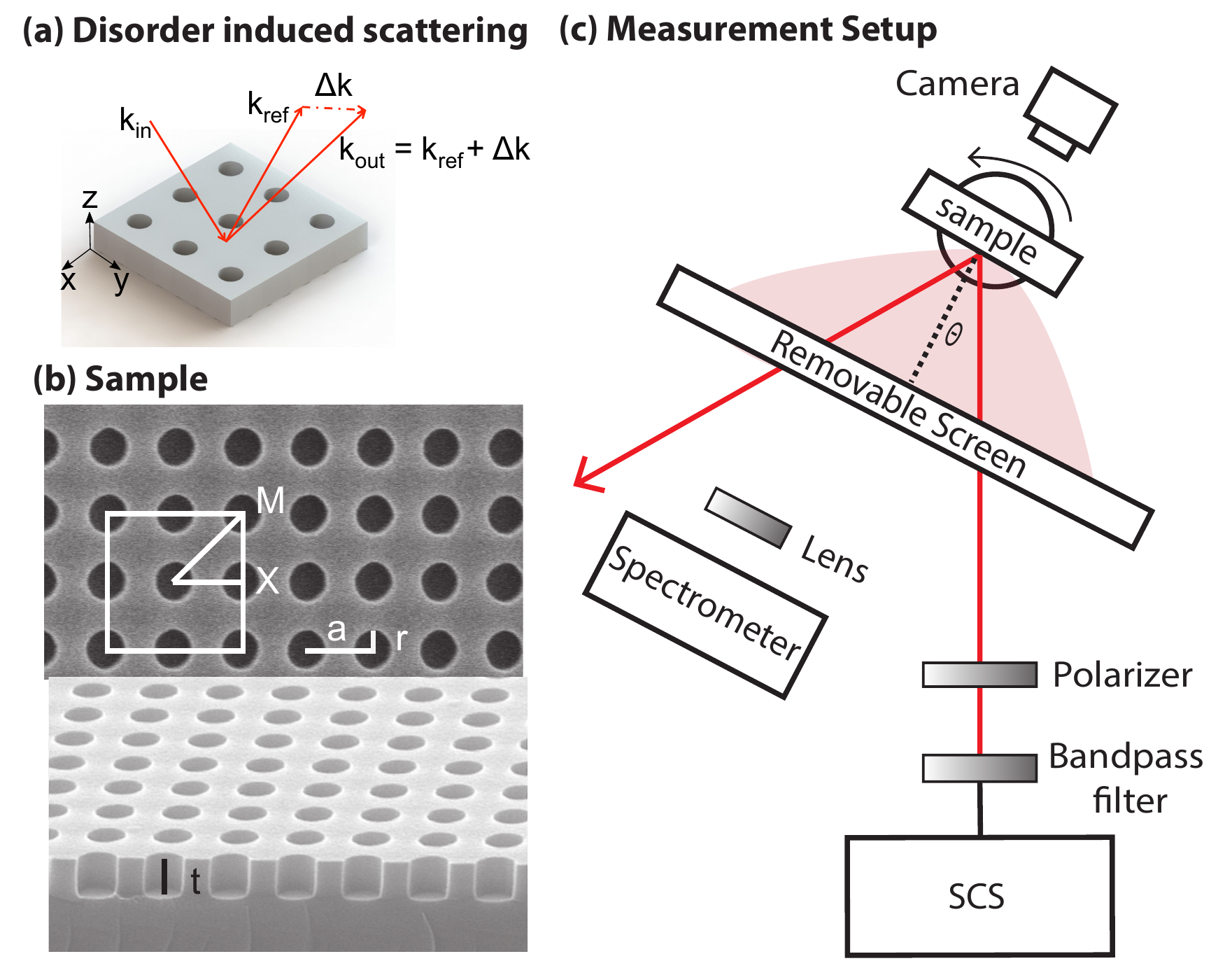}
\caption{\textbf{Resonance-Enhanced Scattering Concept and Experimental Setup.} (a) Scattering of light with incident \textit{in-plane} wavevector $k_{\rm in}$ to wavevector $k_{\rm out}=k_{\rm in}+\Delta k$ due to weak, intrinsic disorder in the sample. (b)  Scanning electron microscope images of the near-pristine PhC samples with $a = 336~\text{nm}$, $r = 103~\text{nm}$, and $t = 180~\text{nm}$: top view (upper panel) and side view (lower panel). (c) Schematic drawing of the experimental setup. The removable screen has slits for the incident and the specularly reflected beams, but it blocks scattered light, showing the projected iso-frequency contours. For a later experiment, the screen and bandpass filter are removed so broadband scattered light can couple into the spectrometer for enhanced-scattering measurements. (SCS: supercontinuum source; SCS and bandpass filter could be replaced with a laser.)}
\label{fig:fig1}
\end{center}
\end{figure}

The direct imaging mechanism relies on a simple process (Fig. \ref{fig:fig1}a). We excite a PhC slab resonance using incident light at the proper frequency $\omega$ and \textit{in-plane} wavevector $(k_\text{x},k_\text{y})$. The natural disorder in the sample scatters light in this resonance to resonances at other wavevectors, depending on the spatial Fourier coefficients of the fabrication disorder. These resonances then radiate photons, which recreates the iso-frequency contour in the far field. By scanning the frequency of the light source and tuning the incident angle accordingly, we can visualize iso-frequency contours throughout the visible spectrum. A quantitative description of the resonance-enhanced scattering process will be presented in the next section. 

\begin{figure*}[htbp]
\begin{center}
\includegraphics[width=6in]{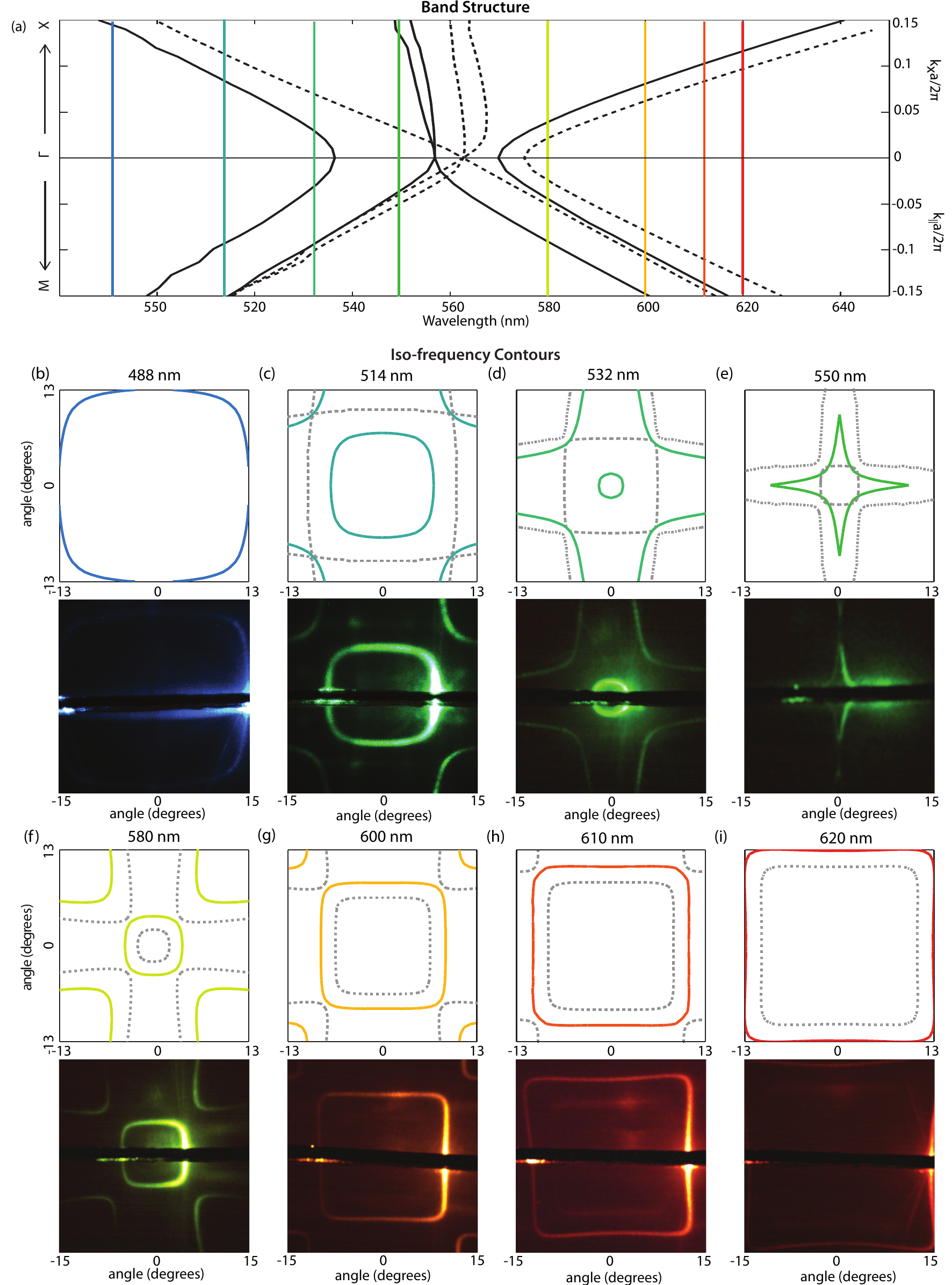}
\caption{\textbf{Direct Visualization of Iso-frequency Contours.} (a) Numerical simulation (COMSOL) of band structure with the colored vertical lines corresponding to wavelengths of the iso-frequency contours below. TE-like bands are dashed and TM-like bands are solid. Numerical (MEEP for b-e, COMSOL for f-i) and experimental iso-frequency contours at (b) $488~\text{nm}$, (c) $514~\text{nm}$, (d) $532~\text{nm}$, (e) $550~\text{nm}$, (f) $580~\text{nm}$, (g) $600~\text{nm}$, (h) $610~\text{nm}$, and (i) $620~\text{nm}$. TE-like contours are dashed and TM-like contours are solid. In all cases, the incident beam excited a TM-like resonance. Experimental data uses incident light with $s-$polarization at angles of (b) $-14.4^{\circ}$, (c) $-8.0^{\circ}$, (d) $-2.6^{\circ}$, (e) $-5.6^{\circ}$, (f) $-5.1^{\circ}$, (g) $-9,8^{\circ}$, (h) $-12.4^{\circ}$, and (i) $-15.4^{\circ}$. The dark, horizontal line in the middle of the experimental contours is the slit in the screen for the incident and the specularly reflected beams to pass through.}
\label{fig:fig2}
\end{center}
\end{figure*}

To demonstrate this technique, we fabricated a hole-array PhC slab and observed the distribution of scattered photons. A large-area PhC slab was fabricated using interference lithography \cite{Lee:14}. As shown in Figure \ref{fig:fig1}b, a high-quality square lattice of cylindrical air holes with radius $103~\text{nm}$ and periodicity $336~\text{nm}$ were patterned into a  $\text{Si}_{3}\text{N}_{4}$ slab of thickness $180~\text{nm}$ on top of a $\text{SiO}_2$ substrate. The sample was then mounted in a  demountable liquid cell filled with liquid of refractive index $n_{liquid}=1.46$, which was then placed on a rotation stage (Newport) for precise control over the incident angle (Fig. \ref{fig:fig1}c). Because we use liquid with the same refractive index as the $\text{SiO}_2$ substrate, our sample has up-down mirror symmetry and thus we can separate the resonances into TM-like (odd-in-$z$) and TE-like (even-in-$z$) modes. The sample was then excited with a broadband supercontinuum source (SuperK, NKT) and a narrow ($10~\text{nm}$) band-pass filter, as shown in Figure \ref{fig:fig1}c, or with a narrow-linewidth laser. A polarizer was placed along the beam path to select either $s$- or $p$-polarization. For this experiment, we chose an $s$-polarized source and chose the incident angle to excite the TM-like resonance. We placed a removable paper screen in front of the sample, with a slit for the incident and reflected beams to pass through. For each wavelength, we used a CMOS camera (Thorlabs DDC1645C) to take an image of the screen that clearly shows the angular distribution of the scattered photons. Examples of photon distribution are shown in Figure \ref{fig:fig2} with excitation wavelengths of $488~\text{nm}$, $514~\text{nm}$, $532~\text{nm}$, $550~\text{nm}$, $580~\text{nm}$, $600~\text{nm}$, $610~\text{nm}$, and $620~\text{nm}$. Contours for $488~\text{nm}$ and $514~\text{nm}$ were created with an argon laser (Modu-laser, Stellar-Pro), and the others were created using the supercontinuum source with bandpass filters. 

We also computed iso-frequency contours for the same frequencies by simulating the band structure $\omega(k_x,k_y)$ using MEEP \cite{OskooiRo10}, a freely-available finite difference time domain (FDTD) software package, and COMSOL, a finite element analysis (FEA) software package. MEEP and COMSOL were used for different frequency ranges to avoid simulation artifacts, which are common when simulating a wide range of angles and frequencies. Our experimental results (Fig. \ref{fig:fig2}, lower panels) show quantitative agreement with the numerical iso-frequency contours for the same wavelengths (Fig. \ref{fig:fig2}, upper panels), with a small, systematic difference in angle present in the experimental setup that can be avoided with a careful angle calibration. 

\section*{Resonance-Enhanced Scattering}

To better understand the imaging technique, we present a theoretical treatment of the resonance-enhanced scattering process. It is well known that macroscopic optical resonances can be used to trap light and therefore enhance non-radiative process, like optical absorption. This leads to intriguing physical phenomena, such as coherent perfect absorption \cite{PhysRevLett.105.053901, Wan18022011}, critically coupled resonators \cite{Tischler:06}, and complete photon absorption in a graphene monolayer \cite{doi:10.1021/ph400090p}. However, the possibility of using large-area resonances to enhance scattering, another type of non-radiative process, remains largely unstudied. In this section, we consider a general PhC slab and compute the enhancement of scattered light from intrinsic fabrication disorder and surface roughness under both on-resonance and off-resonance conditions.

An effective tool to understand this process is the temporal coupled-mode theory (TCMT) \cite{joannopoulos2011photonic,Fan:03,Suh2004}. Consider a collimated beam incident on the PhC slab that excites a resonance. We treat the Rayleigh scattering from the disorder of the slab as a non-radiative decay channel for the resonance. For high-quality PhC slabs made of low-loss dielectrics, the absorptive decay channel is negligible \cite{Lee:14,Lee2012}. When the scattering is weak and lifetime of the resonance is sufficiently long, we can approximate the total amount of scattered light as \cite{joannopoulos2011photonic}:  
\begin{equation}
\label{eq:scatter_TCMT}
\frac{P_{\rm scat}}{P_{\rm in}} = \frac{2\gamma_{\rm r}\gamma_{\rm scat}}{(\omega-\omega_{0})^{2}+(\gamma_{\rm r}+\gamma_{\rm scat})^{2}},
\end{equation}
where $P_{\rm scat}$ is the scattered power, $P_{\rm in}$ is the incident power, $\omega$ is the incoming photon frequency, $\omega_{0}$ is the resonance frequency, $\gamma_{\rm r}$ is the radiative decay rate of the resonance, and $\gamma_{\rm scat}$ is the decay rate due to scattering from disorder (Supplementary Section 1). Here we consider only one resonance in the PhC slab, but the generalization into multiple resonances is straightforward. From equation \eqref{eq:scatter_TCMT}, the scattering rate is most efficient ($P_{\rm scat}/P_{\rm in} = 50\%$) when the system is driven on resonances ($\omega = \omega_{0}$) and the radiative decay rate equals the scattering decay rate ($\gamma_{\rm r} = \gamma_{\rm scat}$). This result agrees with the well-known results in critical coupling for absorption enhancement \cite{Tischler:06, doi:10.1021/ph400090p, PhysRevA.74.064901}. By engineering the quality of the resonances in a photonic crystal, this technique can accommodate samples with different levels of disorder.  

Next, we consider the angular distribution of the scattered photons, which leads to the formation of iso-frequency contours in the far field. The resonance-enhanced scattering can be modeled as a three-step process. In the first step, incoming light with frequency $\omega$ excites a resonance at in-plane wavevector $k_{\rm in}$. In the second step, disorder couples the resonance at $k_{\rm in}$ to resonances at other wavevectors $k_{\rm out} = k_{\rm ref} + \Delta k$. The strength of this coupling depends on the spatial Fourier coefficient of the disorder at $\Delta k$, $|\mathcal{F}_{\Delta k}|^{2}$. In the third step, resonances at $k_{\rm out}$ radiate photons into the far field with outgoing angles specified by frequency $\omega$ and in-plane momentum $k_{\rm out}$. The photons acquire random phases through the disorder coupling (second step), so their subsequent re-radiation (third step) can be modeled as the radiation from a collection of randomly polarized dipoles with number density of $N_{0}$ and dipole strength of $\mu$. As a result, the decay rate into the radiation channel with in-plane wavevector $k_{\text{out}}$ and frequency $\omega$ depends on the spectral density of states (SDOS) \cite{PhysRevE.69.016609} in the vicinity of the PhC slabs and can therefore be approximated as: 
\begin{equation}
\label{eq:SDOS} 
\Gamma(k_{\text{out}},\omega) = \frac{N_{0}\pi \omega |\mu|^{2}}{3\hbar \epsilon} \sum_{n} \alpha_{n} \frac{1}{\pi} \frac{\Delta\omega_{k_{\text{out}}}^{n}}{(\omega-\omega_{k_{\text{out}}}^{n})^{2}+(\Delta\omega_{k_{\text{out}}}^{n})^{2}}.
\end{equation} 
Here, $\mu$ is the electric dipole strength depending on the total power in the scattered photons, $n$ labels different PhC resonances at a given ${k_{\rm out}}$,  $\alpha_{n}$ represents the coupling between the initially excited resonance and the final resonance mediated by the surface roughness, $\Delta\omega_{k_{\text{out}}}^{n}$ are the linewidths of the resonances ($\Delta\omega_{k_{\text{out}}}^{n}=\gamma_{\rm r}+\gamma_{\rm scat}$), and $\omega_{k_{\text{out}}}^{n}$ are the central frequencies of the resonances (Supplementary Section 2). From equation \eqref{eq:SDOS}, it is clear that scattering is maximized when the emission frequency is on-resonance with one of the resonance frequencies for a given momentum ($\omega = \omega_{k_{\text{out}}}^{n}$). Therefore, for incident light with frequency $\omega_{0}$, we expect large scattering for the momenta $k_{\text{out}}$ that provide resonances at $\omega_{0}$. Accordingly, the highest intensity positions in the far field will correspond to the iso-frequency contours of this PhC at the frequency of $\omega_{0}$.

To experimentally demonstrate that this technique relies on resonance-enhanced scattering, we measured the spectrum of the scattered photons from the PhC slab when excited with a supercontinuum source. We removed the paper screen in the experimental setup shown in Figure \ref{fig:fig1}c.  Scattered photons were then collected using a lens with numerical aperture ${\rm NA} = 0.25$ and focal length of $5~\text{cm}$ that was placed in front of a spectrometer with spectral resolution of $0.03~\text{nm}$ (Ocean Optics HR4000). The distance between the spectrometer and the lens was optimized to maximize the coupling of the scattered photons into the spectrometer. While the positions of the spectrometer and the lens were fixed during the experiment, the sample was rotated between $0$ and $8~\text{degrees}$ from the normal direction with a step size of $0.1~\text{degrees}$. Throughout this measurement, we ensured that the directly reflected beam was not captured by the lens, so only scattered light was recorded. 

\begin{figure}[t!]
\begin{center}
\includegraphics[width=3.25in]{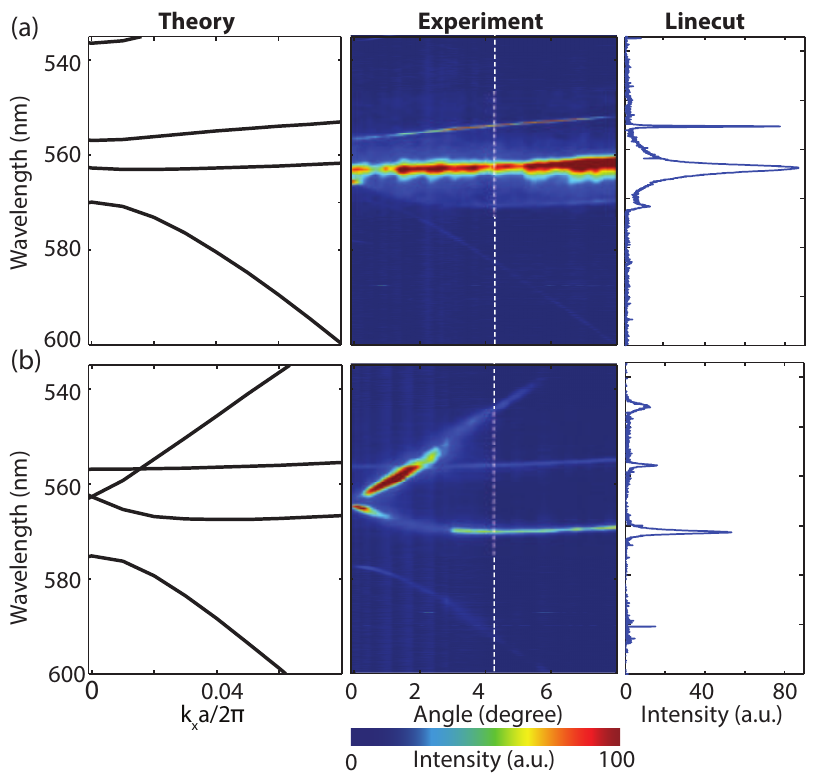}
\caption{\textbf{Experimental Verification of Resonance-Enhanced Scattering.} Numerically calculated band structure (left) and experimentally measured scattering spectrum (middle) for (a) $s$-polarized and (b) $p$-polarized light. The right column shows linecuts at $4.3^{\circ}$ from the experimental scattering spectrum.}
\label{fig:fig3}
\end{center}
\end{figure}

The experimental scattering spectra are shown in the middle column of Figure \ref{fig:fig3}. The scattering peaks show good quantitative agreement with the numerical results for the locations of the resonances obtained from COMSOL (shown in the left column). This can be understood from equation \eqref{eq:scatter_TCMT}: the enhancement is maximized when the incident light frequency is on resonance. $s$-polarized light and $p$-polarized light excite different resonances due to symmetry: the incident beam and the structure are both mirror-symmetric in the $y$ direction. Therefore, the even (odd) incident beam can only excite even (odd) resonances with respect to $y$, as is described in \cite{Lee2012,traplightcont}. Exemplary line-cuts of the scattering spectra at $4.3^{\circ}$ are shown in the right column, which show that the experimental results agree well with the expected Lorentzian lineshapes. 

As we can see in Figure \ref{fig:fig3}, the resonance-enhanced scattering process is also an efficient way to measure PhC band structures, avoiding the shortcomings of other standard techniques. For example, reflection measurements are often used to measure band structures because Fano features in the spectrum indicate resonances \cite{Fan2002}. However, reflection measurements require constantly moving the spectrometer to maintain overlap with the specularly reflected beam. On the other hand, enhanced fluorescence measurements require additional incorporation of emitters and are limited by their emission bandwidth \cite{Zhen20082013}. Finally, enhanced absorption measurements inevitably lower the quality factors of the resonances due to the incorporation of a lossy medium \cite{Xu:09,strasserloss}. The method that we present takes advantage of minimal, generic fabrication errors and surface roughness in any high quality ($Q\approx5,000$) resonator, does not require moving the spectrometer, and allows for a fast and direct measurement of the resonance central frequencies and the quality factors. 

\section*{Discussion}

The experimental iso-frequency contours also give information about the characteristic fabrication errors in the PhC slab, which is not possible to obtain otherwise. As shown in Figure \ref{fig:fig2}, for a given frequency, the contours corresponding to TM-like resonances (solid lines) are brighter than the contours corresponding to the TE-like resonances (dashed lines). This means that the disorder in our sample predominantly couples the initial TM-like resonance to TM-like resonances, which suggests that the major sources of fabrication errors also satisfy up-down mirror symmetry. Therefore, we know that the fabrication errors that do not break up-down mirror symmetry (for example, distortions of the hole shape in the interference process) dominate over  those errors that break the up-down mirror symmetry (for example, oblique side walls in the etching process). Furthermore, we can determine the characteristic length scale of the disorder by analyzing the intensity of the iso-frequency contours (Supplementary Section 3).

Additionally, resonance-enhanced scattering has applications in several display technologies. In particular, transparent displays have been realized using plasmonic resonances \cite{Hsu2014,C5NR06766A}, but the high quality factors of photonic resonances, due to negligible absorption and narrow linewidths, may allow for significantly better performance. If the iso-frequency contour is engineered to be at large angles, and therefore far away from a viewing audience, the resonance-enhanced scattering process can be used for a transparent display. A large portion of the ambient light will pass through the display, but light from a laser projector incident at resonant angles will be scattered and form an image. Projecting incoming light of different polarization states or wavelengths could allow for a transparent 3D display. Further, engineering iso-frequency contours to direct photons to different angles could create a privacy screen \cite{Shen2016}. By projecting light at a specified incident angle, viewers at particular directions could see the strongly-scattered light, while viewers at other angles only see a transparent panel. 

To conclude, this article presents theoretical and experimental results on enhanced photon scattering in the presence of a large-area, nearly pristine PhC slab. We use this phenomenon to measure the band structure of a PhC slab and to directly reconstruct its iso-frequency contours. The iso-frequency contours also give information about the dominant fabrication errors in a sample, and can thus be a useful tool for improving the fabrication process. 

\section*{Acknowledgments}
The authors are grateful to Steve Kooi for his help with the experiments. This work was partly supported by the Army Research Office through the Institute for Soldier Nanotechnologies under contract no. W911NF-13-D-0001. Fabrication of the sample was supported by S3TEC, an Energy Frontier Research Center funded by the US Department of Energy under grant no. DE-SC0001299. B.Z. was partially supported by the United States-Israel Binational Science Foundation (BSF) under award no. 2013508. I.K. was supported in part by Marie Curie grant no. 328853-MC-BSiCS. C.W.H. was partly supported by the National Science Foundation through grant no. DMR-1307632. 

\section*{Author Contributions} 
B.Z. conceived the idea. E.C.R., Y.I., and B.Z. conducted the experiments, and E.C.R. did the numerical analysis. E.C.R. and B.Z. wrote the manuscript, and all other authors edited it.

\bibliographystyle{unsrt}
\bibliography{arxiv_manuscript}

\begin{thebibliography}{10}

\bibitem{PhysRevB.65.201104}
Chiyan Luo, Steven~G. Johnson, J.~D. Joannopoulos, and J.~B. Pendry.
\newblock All-angle negative refraction without negative effective index.
\newblock {\em Phys. Rev. B}, 65:201104, May 2002.

\bibitem{:/content/aip/journal/apl/74/9/10.1063/1.123502}
Hideo Kosaka, Takayuki Kawashima, Akihisa Tomita, Masaya Notomi, Toshiaki
  Tamamura, Takashi Sato, and Shojiro Kawakami.
\newblock Self-collimating phenomena in photonic crystals.
\newblock {\em Applied Physics Letters}, 74(9):1212--1214, 1999.

\bibitem{Rakichsuper}
Peter~T. Rakich, Marcus~S. Dahlem, Sheila Tandon, Mihai Ibanescu, Marin
  Soljacic, Gale~S. Petrich, John~D. Joannopoulos, Leslie~A. Kolodziejski, and
  Erich~P. Ippen.
\newblock Achieving centimetre-scale supercollimation in a large-area
  two-dimensional photonic crystal.
\newblock {\em Nat Mater}, 5(2):93--96, 02 2006.

\bibitem{PhysRevB.58.R10096}
Hideo Kosaka, Takayuki Kawashima, Akihisa Tomita, Masaya Notomi, Toshiaki
  Tamamura, Takashi Sato, and Shojiro Kawakami.
\newblock Superprism phenomena in photonic crystals.
\newblock {\em Phys. Rev. B}, 58:R10096--R10099, Oct 1998.

\bibitem{:/content/aip/journal/apl/97/25/10.1063/1.3524520}
Lei Shi, Haiwei Yin, Xiaolong Zhu, Xiaohan Liu, and Jian Zi.
\newblock Direct observation of iso-frequency contour of surface modes in
  defective photonic crystals in real space.
\newblock {\em Applied Physics Letters}, 97(25), 2010.

\bibitem{PhysRevB.79.033305}
N.~Le~Thomas, R.~Houdr\'e, D.~M. Beggs, and T.~F. Krauss.
\newblock Fourier space imaging of light localization at a photonic band-edge
  located below the light cone.
\newblock {\em Phys. Rev. B}, 79:033305, Jan 2009.

\bibitem{Sapienza2012}
R.~Sapienza, T.~Coenen, J.~Renger, M.~Kuttge, N.~F. van Hulst, and A.~Polman.
\newblock {Deep-subwavelength imaging of the modal dispersion of light}.
\newblock {\em Nature Materials}, 11(9):781--787, 2012.

\bibitem{joannopoulos2011photonic}
J.D. Joannopoulos, S.G. Johnson, J.N. Winn, and R.D. Meade.
\newblock {\em Photonic Crystals: Molding the Flow of Light (Second Edition)}.
\newblock Princeton University Press, 2011.

\bibitem{Fan:03}
Shanhui Fan, Wonjoo Suh, and J.~D. Joannopoulos.
\newblock Temporal coupled-mode theory for the fano resonance in optical
  resonators.
\newblock {\em J. Opt. Soc. Am. A}, 20(3):569--572, Mar 2003.

\bibitem{Suh2004}
Wonjoo Suh, Zheng Wang, and Shanhui Fan.
\newblock {Temporal coupled-mode theory and the presence of non-orthogonal
  modes in lossless multimode cavities}.
\newblock {\em IEEE Journal of Quantum Electronics}, 40(10):1511--1518, 2004.

\bibitem{McCutcheon2005}
Murray~W. McCutcheon, Georg~W. Rieger, Iva~W. Cheung, Jeff~F. Young, Dan
  Dalacu, Simon Frederick, Philip~J. Poole, Geof~C. Aers, and Robin~L.
  Williams.
\newblock {Resonant scattering and second-harmonic spectroscopy of planar
  photonic crystal microcavities}.
\newblock {\em Applied Physics Letters}, 87(22):1--3, 2005.

\bibitem{Galli2009}
M.~Galli, S.~L. Portalupi, M.~Belotti, L.~C. Andreani, L.~O'Faolain, and T.~F.
  Krauss.
\newblock {Light scattering and Fano resonances in high-Q photonic crystal
  nanocavities}.
\newblock {\em Applied Physics Letters}, 94(7):1--3, 2009.

\bibitem{Portalupi2011}
S.~L. Portalupi, M.~Galli, M.~Belotti, L.~C. Andreani, T.~F. Krauss, and
  L.~O'faolain.
\newblock {Deliberate versus intrinsic disorder in photonic crystal
  nanocavities investigated by resonant light scattering}.
\newblock {\em Physical Review B - Condensed Matter and Materials Physics},
  84(4):1--9, 2011.

\bibitem{Lee:14}
Jeongwon Lee, Bo~Zhen, Song-Liang Chua, Ofer Shapira, and Marin
  Solja\v{c}i\'{c}.
\newblock Fabricating centimeter-scale high quality factor two-dimensional
  periodic photonic crystal slabs.
\newblock {\em Opt. Express}, 22(3):3724--3731, Feb 2014.

\bibitem{OskooiRo10}
Ardavan~F. Oskooi, David Roundy, Mihai Ibanescu, Peter Bermel, J.~D.
  Joannopoulos, and Steven~G. Johnson.
\newblock {MEEP}: A flexible free-software package for electromagnetic
  simulations by the {FDTD} method.
\newblock {\em Comput. Phys. Commun.}, 181:687--702, January 2010.

\bibitem{PhysRevLett.105.053901}
Y.~D. Chong, Li~Ge, Hui Cao, and A.~D. Stone.
\newblock Coherent perfect absorbers: Time-reversed lasers.
\newblock {\em Phys. Rev. Lett.}, 105:053901, Jul 2010.

\bibitem{Wan18022011}
Wenjie Wan, Yidong Chong, Li~Ge, Heeso Noh, A.~Douglas Stone, and Hui Cao.
\newblock Time-reversed lasing and interferometric control of absorption.
\newblock {\em Science}, 331(6019):889--892, 2011.

\bibitem{Tischler:06}
Jonathan~R. Tischler, M.~Scott Bradley, and Vladimir Bulovi\'{c}.
\newblock Critically coupled resonators in vertical geometry using a planar
  mirror and a 5 nm thick absorbing film.
\newblock {\em Opt. Lett.}, 31(13):2045--2047, Jul 2006.

\bibitem{doi:10.1021/ph400090p}
Jessica~R. Piper and Shanhui Fan.
\newblock Total absorption in a graphene monolayer in the optical regime by
  critical coupling with a photonic crystal guided resonance.
\newblock {\em ACS Photonics}, 1(4):347--353, 2014.

\bibitem{Lee2012}
Jeongwon Lee, Bo~Zhen, Song~Liang Chua, Wenjun Qiu, John~D. Joannopoulos, Marin
  Solja\v{c}i\'{c}, and Ofer Shapira.
\newblock {Observation and differentiation of unique high-Q optical resonances
  near zero wave vector in macroscopic photonic crystal slabs}.
\newblock {\em Physical Review Letters}, 109(6):1--5, 2012.

\bibitem{PhysRevA.74.064901}
David L.~C. Chan, Ivan Celanovic, J.~D. Joannopoulos, and Marin Solja\ifmmode
  \check{c}\else \v{c}\fi{}i\ifmmode~\acute{c}\else \'{c}\fi{}.
\newblock Emulating one-dimensional resonant $q$-matching behavior in a
  two-dimensional system via fano resonances.
\newblock {\em Phys. Rev. A}, 74:064901, Dec 2006.

\bibitem{PhysRevE.69.016609}
R.~C. McPhedran, L.~C. Botten, J.~McOrist, A.~A. Asatryan, C.~M. de~Sterke, and
  N.~A. Nicorovici.
\newblock Density of states functions for photonic crystals.
\newblock {\em Phys. Rev. E}, 69:016609, Jan 2004.

\bibitem{traplightcont}
Chia~Wei Hsu, Bo~Zhen, Jeongwon Lee, Song-Liang Chua, Steven~G. Johnson,
  John~D. Joannopoulos, and Marin Soljacic.
\newblock Observation of trapped light within the radiation continuum.
\newblock {\em Nature}, 499(7457):188--191, 07 2013.

\bibitem{Fan2002}
Shanhui Fan and J.~Joannopoulos.
\newblock {Analysis of guided resonances in photonic crystal slabs}.
\newblock {\em Physical Review B}, 65(23):235112, June 2002.

\bibitem{Zhen20082013}
Bo~Zhen, Song-Liang Chua, Jeongwon Lee, Alejandro~W. Rodriguez, Xiangdong
  Liang, Steven~G. Johnson, John~D. Joannopoulos, Marin Soljacic, and Ofer
  Shapira.
\newblock Enabling enhanced emission and low-threshold lasing of organic
  molecules using special fano resonances of macroscopic photonic crystals.
\newblock {\em Proceedings of the National Academy of Sciences},
  110(34):13711--13716, 2013.

\bibitem{Xu:09}
Tao Xu, Mark~S. Wheeler, Harry~E. Ruda, Mohammad Mojahedi, and J.~Stewart
  Aitchison.
\newblock The influence of material absorption on the quality factor of
  photonic crystal cavities.
\newblock {\em Opt. Express}, 17(10):8343--8348, May 2009.

\bibitem{strasserloss}
Roman Gansch, Stefan Kalchmair, Patrice Genevet, Tobias Zederbauer, Hermann
  Detz, Aaron~M. Andrews, Werner Schrenk, Federico Capasso, Marko Loncar, and
  Gottfried Strasser.
\newblock Measurement of bound states in the continuum by a detector embedded
  in a photonic crystal.
\newblock {\em Light: Science and Applications}.

\bibitem{Hsu2014}
Chia~Wei Hsu, Bo~Zhen, Wenjun Qiu, Ofer Shapira, Brendan~G DeLacy, John~D
  Joannopoulos, and Marin Solja\v{c}i\'{c}.
\newblock {Transparent displays enabled by resonant nanoparticle scattering}.
\newblock {\em Nat Commun}, 5, January 2014.

\bibitem{C5NR06766A}
Koichiro Saito and Tetsu Tatsuma.
\newblock A transparent projection screen based on plasmonic {A}g nanocubes.
\newblock {\em Nanoscale}, 7:20365--20368, 2015.

\bibitem{Shen2016}
Yichen Shen, Chia~Wei Hsu, Yi~Xiang Yeng, John~D. Joannopoulos, and Marin
  Solja\v{c}i\'{c}.
\newblock {Broadband angular selectivity of light at the nanoscale: Progress,
  applications, and outlook}.
\newblock {\em Applied Physics Reviews}, 3(1), 2016.

\bibitem{Zhen2013}
Bo~Zhen, Song-Liang Chua, Jeongwon Lee, Alejandro~W Rodriguez, Xiangdong Liang,
  Steven~G Johnson, John~D Joannopoulos, Marin Soljacic, and Ofer Shapira.
\newblock {Enabling enhanced emission and low-threshold lasing of organic
  molecules using special Fano resonances of macroscopic photonic crystals.}
\newblock {\em Proceedings of the National Academy of Sciences of the United
  States of America}, 110(34):13711--6, 2013.

\bibitem{principlesofnanooptics}
L.~Novotny and B.~Hecht.
\newblock {\em Principles of Nano-Optics}.
\newblock Cambridge University Press, 2006.

\bibitem{803000}
M.~Boroditsky, R.~Vrijen, T.~F. Krauss, R.~Coccioli, R.~Bhat, and
  E.~Yablonovitch.
\newblock Spontaneous emission extraction and purcell enhancement from
  thin-film 2-d photonic crystals.
\newblock {\em Journal of Lightwave Technology}, 17(11):2096--2112, Nov 1999.

\end{thebibliography}

\section*{Supplementary}

\renewcommand{\theequation}{S\arabic{equation}}
\setcounter{equation}{0}

\renewcommand{\thefigure}{S\arabic{figure}}
\setcounter{figure}{0}

\noindent {\bf Sample Fabrication:} The $\text{Si$_{3}$N$_{4}$}$ layer was grown by low-pressure chemical vapor deposition on a $6\text{$\mu$m}$-thick cladding of $\text{SiO$_2$}$ on the backbone of a silicon wafer (LioniX). The wafer was then coated with an anti-reflection polymer coating, a thin $\text{SiO$_2$}$ intermediate layer, and a layer of negative photoresist. The square lattice pattern was created with Mach Zehnder interference lithography using a 325-nm He/Cd laser. The angle between the two arms of the laser beam was chosen for a periodicity of 336 nm, and the hole exposure was selected for a hole radius of $103~\text{nm}$. After exposures, the pattern in the photoresist was transferred to the $\text{Si$_{3}$N$_{4}$}$ by reactive-ion etching.

\noindent {\bf Derivation of Total Scattered Light:}
 \begin{figure}[b!]
 \begin{center}
 \includegraphics[width=3.25in]{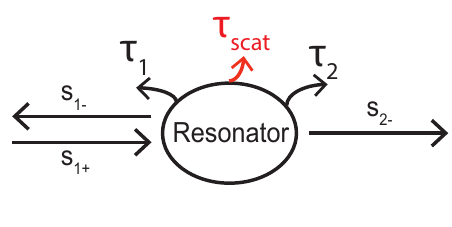}
 \caption{\textbf{Temporal Coupled-Mode Theory Schematic.} Schematic of a resonator with input channel $s_{1+}$, output channels $s_{1-}$ and $s_{2-}$, and resonant decay channels with lifetimes $\tau_1$, $\tau_2$, and $\tau_{scat}$.}
 \label{fig:tcmt}
 \end{center}
 \end{figure}
 We determine the total scattered power using temporal coupled-mode theory \cite{joannopoulos2011photonic,Fan:03,Suh2004}. The system consists of a PhC slab resonator with one input port and two output ports (Fig. \ref{fig:tcmt}). Let the incoming wave $s_{1+}$ couple to the resonator with coupling constant $\kappa$, and let the resonant mode couple to the outgoing waves $s_{1-}$ and $s_{2-}$ with coupling constants $d_1$ and $d_2$, respectively. In addition to the resonance-assisted processes, the incoming and outgoing waves are coupled by a direct scattering matrix $C$, which has the form
\begin{equation}
C = \begin{pmatrix} r & -it \\ -it & r \end{pmatrix},
\label{tcmtC}
\end{equation}
where $r$ and $t$ are constants with $|r|^2+|t|^2=1$.  In this case, the direct reflection is small because of the low index contrast between the slab and the surrounding liquid, so we can approximate $r=0$ and $t=i$. Let $A$ be the amplitude of the resonant electromagnetic field, which is normalized such that $|A|^2$ is the energy in the slab. With an incident field oscillating at  frequency $\omega$, $A$ can be written as
\begin{equation}
\frac{dA}{dt}=-i\omega A=\left(-i\omega_0-\frac{1}{\tau}\right)A+\kappa s_{1+}
\label{tcmt1}
\end{equation}
\begin{equation}
\begin{pmatrix} s_{1-} \\ s_{2-} \end{pmatrix}= \begin{pmatrix} 0 & 1 \\ 1 & 0 \end{pmatrix} \begin{pmatrix} s_{1+} \\ 0 \end{pmatrix} + A\begin{pmatrix} d_{1} \\ d_{2} \end{pmatrix}.
\label{tcmt2}
\end{equation}
Here, $\omega_0$ is the center frequency of the resonance and $\tau$ is the lifetime of the resonance. $\tau$ can be broken down into several decay channels [Fig. \ref{fig:tcmt}]. First, the resonance can decay into two radiative channels, with lifetimes $\tau_1$ and $\tau_2$. Additionally, the resonance can decay into a nonradiative scattering channel with lifetime $\tau_{scat}$. For high-quality PhC slabs made of low-loss dielectrics, absorption is negligible, so we do not need to consider an absorption decay channel. Then $1/\tau=1/\tau_{1}+1/\tau_{2}+1/\tau_{scat}$.
  
Conservation of energy requires that $|d_1|^2=2/\tau_1$ and $|d_2|^2=2/\tau_2$. Then because of the symmetry of the system, $\tau_1=\tau_2$. Finally, we excite odd modes with respect to $z$, so $d_1=-d_2$. Defining the radiative lifetime $1/\tau_{r}=1/\tau_1 + 1/\tau_2$, this gives
\begin{equation}
\begin{pmatrix} d_{1} \\ d_{2} \end{pmatrix}=\begin{pmatrix} \sqrt{1/\tau_r} \\ -\sqrt{1/\tau_r} \end{pmatrix}.
\label{tcmt3}
\end{equation}
Because of time-reversal symmetry, $\kappa=d_1$. For simplicity, we can define $\gamma_i$ as the decay rate for each channel, such that $\gamma_i=1/\tau_i$. Then equations \ref{tcmt1} and \ref{tcmt2} become: 
\begin{equation}
\frac{dA}{dt}=\left(-i\omega_0-\gamma_r-\gamma_{scat}\right)A+\sqrt{\gamma_r} s_{1+}
\label{tcmt52}
\end{equation}
\begin{equation}
\begin{pmatrix} s_{1-} \\ s_{2-} \end{pmatrix}= \begin{pmatrix} 0 & 1 \\ 1 & 0 \end{pmatrix} \begin{pmatrix} s_{1+} \\ 0 \end{pmatrix} + A\begin{pmatrix} \sqrt{\gamma_r} \\ -\sqrt{\gamma_r} \end{pmatrix}.
\label{tcmt62}
\end{equation}
The scattered power is given by $1-R-T$, where $R$ and $T$ are reflection coefficients given below. 
\begin{equation}
R = \frac{|s_{1-}|^2}{|s_{1+}|^2} = \frac{\gamma_r^2}{(\omega-\omega_0)^2+(\gamma_r+\gamma_{scat})^2}
\label{tcmtR}
\end{equation}
\begin{equation}
T = \frac{|s_{2-}|^2}{|s_{1+}|^2} = \frac{(\omega-\omega_0)^2+\gamma_{scat}^2}{(\omega-\omega_0)^2+(\gamma_r+\gamma_{scat})^2}
\label{tcmtT}
\end{equation}
The scattered power is then
\begin{equation}
\begin{split}
\frac{P_{scat}}{P_{incident}} &= 1-T-R \\
&= \frac{2\gamma_r \gamma_{scat}}{(\omega-\omega_0)^2+(\gamma_r+\gamma_{scat})^2}.
\label{tcmtSCAT}
\end{split}
\end{equation}

\noindent {\bf Derivation of Resonance Decay Rate:} The following is a simple derivation to highlight the physical process; a more rigorous version involves decomposing the Green's function of the system in the basis of normalized Bloch modes \cite{Zhen2013}. Consider an electric dipole in free space. The free-space spontaneous decay rate of the dipole is \cite{principlesofnanooptics}
\begin{equation}
\Gamma_0 = \frac{\omega_0^3|\boldsymbol{\mu}|^2}{3\pi \epsilon_0 \hbar c^3},
\label{dipole1}
\end{equation}
where $\boldsymbol{\mu}$ is the dipole moment. The decay rate is proportional to the density of states $D(\omega)$ of the system. In free space, the number of eigenmodes in a volume $V$ with eigenfrequencies less than $\omega_0$ is 
\begin{equation}
\begin{split}
N(\omega_0) & = 2\times \frac{V}{(2\pi)^3}\int_{k<\omega_0/c}dk\\
& = \frac{\omega_0^3 V}{3\pi^2c^3}, \\
\end{split}
\label{N_DOS}
\end{equation}
so the density of states is 
\begin{equation}
\begin{split}
D(\omega) & = \frac{\partial N(\omega)}{\partial \omega} \\
& = \frac{\omega^2 V}{\pi^2 c^3}. \\
\end{split}
\label{DOS}
\end{equation}
If we divide equation \ref{dipole1} by $D(\omega),$ we find 
\begin{equation}
\begin{split}
\frac{\Gamma_0}{D(\omega)} & =\frac{\pi\omega|\boldsymbol{\mu}|^2}{V3\hbar \epsilon_0}\\
& = \frac{N_0 \pi\omega|\boldsymbol{\mu}|^2}{3\hbar \epsilon_0}
\end{split},
\label{dipole2}
\end{equation}
where $N_0$ is the number density of dipoles. We then multiply by the spectral density of states of the photonic crystal, which we can express as a sum of Lorentzian for each resonance, as in Ref. \cite{803000}.

\begin{equation}
\label{decayrate} 
\Gamma(k_{\text{out}},\omega) = \frac{N_{0}\pi \omega |\mu|^{2}}{3\hbar \epsilon} \sum_{n} \alpha_{n} \frac{1}{\pi} \frac{\Delta\omega_{k_{\text{out}}}^{n}}{(\omega-\omega_{k_{\text{out}}}^{n})^{2}+(\Delta\omega_{k_{\text{out}}}^{n})^{2}}.
\end{equation} 
where $n$ labels different PhC resonances at a given $k_{\rm out}$,  $\alpha_{n}$ represents the coupling between the initially excited resonance and the final resonance mediated by the surface roughness, $\Delta\omega_{k_{\text{out}}}^{n}$ are the linewidths of the resonances ($\Delta\omega_{k_{\text{out}}}^{n}=\gamma_{\rm r}+\gamma_{\rm scat}$), and $\omega_{k_{\text{out}}}^{n}$ are central frequencies of the resonances.

\begin{figure}[b!]
 \begin{center}
 \includegraphics[width=3.25in]{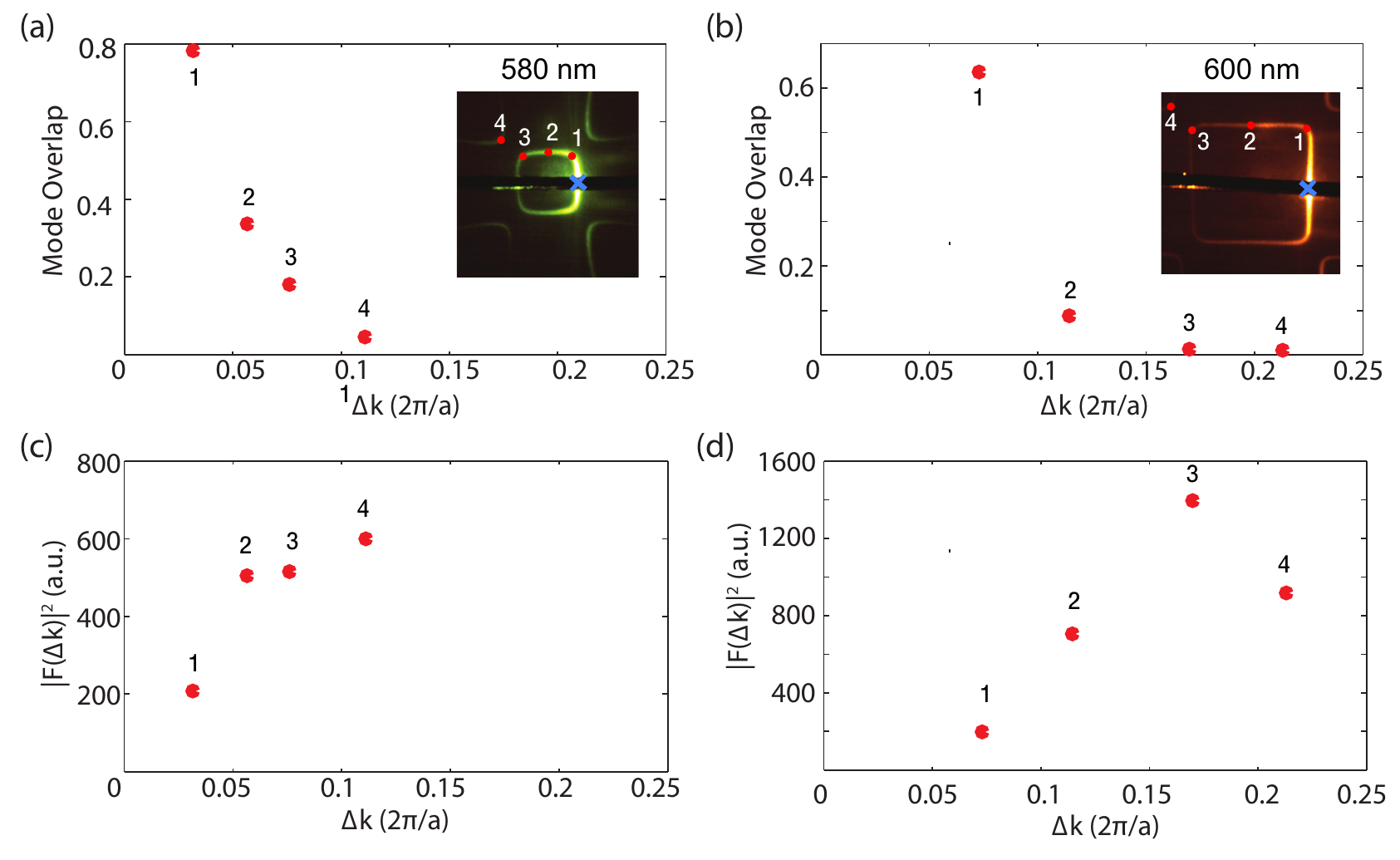}
 \caption{\textbf{Spatial Characterization of Disorder.} The top row shows numerical (FDTD) simulations of the mode overlap between resonant electric fields at \textit{in-plane} wavevectors $k_{\text{in}}$ and $k_{\text{out}}=k_{\text{ref}}+\Delta k$ for four points in the (a)  $580~\text{nm}$ and (b) $600~\text{nm}$ contours. Insets: the blue cross corresponds to the reflected beam, and numbered red dots refer to the scattered light at four $k$-points studied here. The bottom row shows calculated spatial Fourier coefficients of the disorder for the same four points in (c) $580~\text{nm}$ and (d) $600~\text{nm}$ contours. The $580~\text{nm}$ contours have a smaller spread in $k$, which gives data on the lower end of $\Delta k$ that is complimented by the data from the $600~\text{nm}$ range.}
 \label{fig:disorder}
 \end{center}
 \end{figure}

\noindent {\bf Disorder Characterization: } We note that the intensity of the iso-frequency contours drops off as the change of momentum $\Delta k$ increases. The intensity $I$ at each point in the contour depends on two factors: the spatial Fourier coefficient of the disorder $F(\left|\Delta k\right|)$ and the spatial mode overlap between the initially excited resonance and the final resonance. 
 \begin{equation}
 \label{disorder}
 I(k_x,k_y)\propto|F(\left|\Delta k\right|)|^2\times\left|\left\langle \textbf{u}_{k_\text{in}}(\textbf{r})|\epsilon(\textbf{r}) \textbf{u}_{k_\text{out}}(\textbf{r})\right\rangle\right|^2,
 \end{equation}
where $\textbf{u}_{k_\text{in}}$ and $\textbf{u}_{k_\text{out}}$ are the normalized electric fields of the initially excited resonance and scattered resonance, respectively. Therefore, using the intensity in the experimental contours, we can determine the characteristic Fourier coefficients of the disorder by computing the mode overlap using FDTD methods \cite{OskooiRo10}. As expected, the results for $580~\text{nm}$ and $600~\text{nm}$ contours indicate that the spatial overlap between incoming and outgoing resonances decreases greatly as $\Delta k$ increases (Fig. \ref{fig:disorder}a-b). However, the intensity of the iso-frequency contours drops off at a slower rate, so the spatial Fourier coefficient of the disorder increases with $\Delta k$ (Fig. \ref{fig:disorder}c-d). The peak in the Fourier coefficient indicates that the characteristic length scale of the disorder is on the order of a few unit cells. A more complete statistical analysis and comparison with fabricated structures is required to understand the full potential of this technique.  

\end{document}